\documentclass[final,3p,times,twocolumn]{elsarticle}
 \biboptions{comma,sort&compress}
 \usepackage{graphicx}
  \usepackage{epsfig}

\def\nin{\noindent}
\def\be{\begin{equation}}
\def\ee{\end{equation}}
\def\bea{\begin{eqnarray}}
\def\eea{\end{eqnarray}}
\def\nn{\nonumber}

\journal{Nuc. Phys. (Proc. Suppl.)}

\begin{document}

\begin{frontmatter}



\title{$B_s \to f_0(980)$ decays: Results from light-cone QCD Sum Rules }

 \author[label1]{Fulvia De Fazio}
  \address[label1]{Istituto Nazionale di Fisica Nucleare INFN - Sezione di Bari,
\\
Via Orabona 4, I-70126 - Bari, Italy.}
\ead{fulvia.defazio@ba.infn.it}

\begin{abstract}
\noindent
We describe a light-cone
QCD sum rule calculation of the $B_s\to f_0(980)$ transition form factors  useful to predict the branching ratios of the rare
decays $B_s \to f_0 \ell^+ \ell^-$, $B_s \to f_0 \nu \bar \nu$ and  of $B_s\to J/\psi
f_0$ decay   assuming factorization. We   compare  this channel to  $B_s \to J/\psi \phi$ as far as the possibility to  determine  the $B_s$ mixing phase is concerned.
\end{abstract}


\end{frontmatter}


\section{Introduction}
\nin
Rare $B_s$ decays induced  at loop level in the Standard Model (SM) are sensitive to new physics (NP)  effects that may enhance their  small  branching ratios \cite{Ball:2000ba}.
Besides, the analysis of the $B_s$ unitarity triangle of  Cabibbo-Kobayashi-Maskawa (CKM) elements: $ V_{us}V_{ub}^*  + V_{cs}V_{cb}^* + V_{ts}V_{tb}^*  = 0 $
provides  an important test of the SM description of CP violation.
One of its angles, $ \beta_s= Arg\left[ - \frac{V_{ts} V^*_{tb}}{V_{cs} V^*_{cb}}\right]$,  is expected  to be tiny in the SM: $\beta_s \simeq 0.019$ rad.
Recently CDF~\cite{Aaltonen:2007he} and D0~\cite{:2008fj} Collaborations have indicated larger values with sizable uncertainties, although the latest CDF analysis \cite{oakes} seems to reconcile the SM with data.
 Hence the precise measurement of $\beta_s$ is a priority for forthcoming experiments.
\\ \nin In this paper we describe the light-cone QCD sum rule (LCSR) calculation of the $B_s \to f_0(980)$~\footnote{Hereafter, we
use $f_0$ to denote the $f_0(980)$ meson.} form factors  \cite{Colangelo:2010bg},
using the results to predict the branching ratios of the  decays $B_s\to f_0 \ell^+ \ell^-$, $B_s\to f_0\nu\bar\nu$ in the SM. We also study the  mode  $B_s\to J/\psi
f_0$ that allows to access  $\beta_s$ \cite{Stone:2008ak}.
%
\section{ $B_s\to f_0$ form factors in Light-Cone  Sum Rules \label{sec:formfactors-LCSR}}
\nin
The matrix elements involved in $B_s\to f_0$ transitions can be parameterized  in terms of  form factors as
\begin{eqnarray}
\hspace{-0.6cm} \langle f_0(p_{f_0})|\bar s \gamma_\mu\gamma_5 b |\overline {B}_s(p_{B_s})\rangle=&&   \label{F1-F0} \\
\hspace{-0.6cm}-i \Big\{F_1(q^2)\Big[P_\mu
 -\frac{m_{B_s}^2-m_{f_0}^2}{q^2}q_\mu\Big]\hspace{-0.2cm} &+& \hspace{-0.2cm}F_0(q^2)\frac{m_{B_s}^2-m_{f_0}^2}{q^2}q_\mu\Big\}, \,\,\, \nonumber
\eea
 \bea
 &&\langle{f_0}(p_{f_0})|\bar s\sigma_{\mu\nu}\gamma_5q^\nu b |\overline {B}_s(p_{B_s})\rangle = \nonumber \\
&&-\frac{F_T(q^2)}{m_{B_s}+m_{f_0}}   \Big[q^2P_\mu
 -(m_{B_s}^2-m_{f_0}^2)q_\mu\Big], \label{FT}
\end{eqnarray}
where $P=p_{B_s}+p_{f_0}$ and $q=p_{B_s}-p_{f_0}$.
To compute such form factors
 using light-cone QCD sum rules (LCSR) \cite{Colangelo:2000dp} we consider the
correlation function:
\begin{eqnarray}
 \hspace{-0.7cm} \Pi(p_{f_0},q)= i \int d^4x \, e^{iq\cdot x} \langle {f_0}(p_{f_0})|{\rm
 T}\left\{j_{\Gamma_1}(x),j_{\Gamma_2}(0)\right\}|0\rangle \hspace{-0.1cm}
 \label{corr}
\end{eqnarray}
where $j_{\Gamma_1}$ is one of the currents
in the  definitions (\ref{F1-F0})-(\ref{FT})  of
the form factors: $j_{\Gamma_1}=J_\mu^5=\bar s\gamma_\mu\gamma_5b$
for $F_1$ and $F_0$, and $j_{\Gamma_1}=J_\mu^{5T}=\bar
s\sigma_{\mu\nu}\gamma_5 q^\nu b$ for $F_T$.   $j_{\Gamma_2}=\bar b i\gamma_5 s$ interpolates  the $B_s$ meson;
its matrix element between the vacuum and $B_s$ defines the  $f_{B_s}$ decay
constant:
$\langle \overline B_s(p_{B_s})| \bar b i\gamma_5 s|0\rangle =
 \frac{m_{B_s}^2}{m_{b}+m_s}f_{B_s}$.
The LCSR method consists in evaluating the correlator
(\ref{corr}) both at the hadronic  level and in QCD. Equating
the two  representations gives a sum rule suitable  to derive  the form
factors.

The hadronic representation of the correlator in
(\ref{corr}) can be written as
  the contribution of the $\bar B_s$ plus that of the
higher resonances and the continuum of states $h$:
\begin{eqnarray}
 \hspace{-1.4cm}&&\Pi^{\rm H}(p_{f_0},q)=  \frac{\langle {f_0}(p_{f_0})|j_{\Gamma_1}|\overline
 {B}_s( p_{f_0}+q)\rangle \langle \overline {B}_s(p_{f_0}+q)|j_{\Gamma_2}|0\rangle}
 {m_{B_s}^2-(p_{f_0}+q)^2}\nonumber \\\hspace{-1.4cm}&&+ \int_{s_0}^\infty ds \frac{\rho^h(s,q^2)}
 {s-(p_{f_0}+q)^2}, \label{hadronic}
\end{eqnarray}
where  higher resonances and  the continuum of states are
described in terms of the spectral function $\rho^h(s,q^2)$, which
 contributes starting from a threshold $s_0$.

\nin To evaluate  the correlator  in QCD we write   it  as
\begin{eqnarray}
 \Pi^{\rm QCD}(p_{f_0},q)=   \frac{1}{\pi}\int_{(m_b+m_s)^2}^\infty ds \, \frac{{\rm Im}\Pi^{\rm QCD}(s,q^2)}
 {s-(p_{f_0}+q)^2} \,. \label{QCD-repr}
\end{eqnarray}
Expanding the T-product in  (\ref{corr}) on the light-cone, we obtain a series of operators, ordered by increasing twist, the matrix elements of which between the vacuum and the $f_0$  are written in terms of  $f_0$ light-cone distribution amplitues (LCDA).
Since the  function $\rho^h$ in (\ref{hadronic}) is unknown, we use global quark-hadron duality
to identify $\rho^h$ with $\rho^{\rm QCD}={1 \over \pi} {\rm Im} \Pi^{\rm QCD}$ when integrated
above $s_0$  \cite{shifman-duality}:
\begin{eqnarray}
\vspace{-0.8cm}\int_{s_0}^\infty  ds {\rho^h(s,q^2) \over s-(p_{f_0}+q)^2}=\frac{1}{\pi}\int_{s_0}^\infty ds \, \frac{{\rm Im}\Pi^{\rm QCD}(s,q^2)}{s-(p_{f_0}+q)^2}\,. \nonumber
\end{eqnarray}
Using duality, together with  the equality $\Pi^{\rm H}(p_{f_0},q)=\Pi^{\rm QCD}(p_{f_0},q)$,
we obtain from Eqs. (\ref{hadronic}) and (\ref{QCD-repr}):
\begin{eqnarray}\hspace{-0.3cm}&&
  \frac{\langle {f_0}(p_{f_0})|j_{\Gamma_1}|\overline
 {B_s}( p_{f_0}+q)\rangle \langle \overline {B_s}(p_{f_0}+q)|j_{\Gamma_2}|0\rangle}
 {m_{B_s}^2-(p_{f_0}+q)^2}=\nonumber \\ \hspace{-0.3cm}&&\frac{1}{\pi}\int_{(m_b+m_s)^2}^{s_0} ds \, \frac{{\rm Im}\Pi^{\rm QCD}(s,q^2)} {s-(p_{f_0}+q)^2} \,\,\ .
\label{res1}
\end{eqnarray}
We perform a Borel
 transformation of  the two sides in (\ref{res1}),  exploiting the result
$\displaystyle{{\cal B} \left[ { 1 \over (s+Q^2)^n } \right]={\exp(-s/M^2) \over (M^2)^n\ (n-1)!}}$  where $Q^2=-q^2$ and $M^2$ is  the Borel parameter.
This operation improves the convergence of the series in $\Pi^{\rm QCD}$  and for suitable values of $M^2$
enhances the contribution of the low lying states to  $\Pi^{\rm H}$.
Applying it to  $\Pi^{\rm H}$ and $\Pi^{\rm QCD}$ we
get
\begin{eqnarray}
\hspace{-1cm}  &&  {\langle {f_0}(p_{f_0})|j_{\Gamma_1}|\overline {B}_s(p_{B_s})\rangle \langle \overline {B}_s(p_{B_s})|j_{\Gamma_2}|0\rangle}
 {\rm
 exp}\left[-\frac{m_{B_s}^2}{M^2}\right]=\nonumber \\\hspace{-1cm}
 &&\frac{1}{\pi}\int_{(m_b+m_s)^2}^{s_0}ds \,\,
 {\rm exp}[-s/M^2] \, \, {\rm Im}\Pi^{\rm QCD}(s,q^2) \label{SR-generic}
\end{eqnarray}
($p_{B_s}=p_{f_0}+q$). From (\ref{SR-generic})  we
derive the sum rules for $F_1$, $F_0$ and $F_T$,  choosing
either   $j_{\Gamma_1}=J_\mu^5$ or   $j_{\Gamma_1}=J_\mu^{5T}$.

In the calculation of $\Pi^{\rm QCD}$ we consider $f_0$ as a $s \bar s$ state modified by some hadronic dressing \cite{DeFazio:2001uc}.
Possible $f_0-\sigma$ mixing \cite{f0sigma}
 may  only affect the overall normalization of the form factors at zero recoil, a systematic  uncertainty  in our numerical results.
We refer to \cite{Colangelo:2010bg} for the  definitions of the $f_0$ LCDA, for numerical values of the input parameters as well as for the final expressions of the form factors obtained from (\ref{SR-generic}).
 We  fix
 $s_0=(34\pm 2)\, {\rm
 GeV}^2$, which should correspond to the mass squared of the first radial excitation of $B_s$.
 As for the Borel parameter,  the form factors for
each value of $q^2$ depend on it. The
result is obtained requiring stability against variations of
$M^2$.
\begin{figure}[b]\begin{center}
\includegraphics[width=0.35\textwidth]{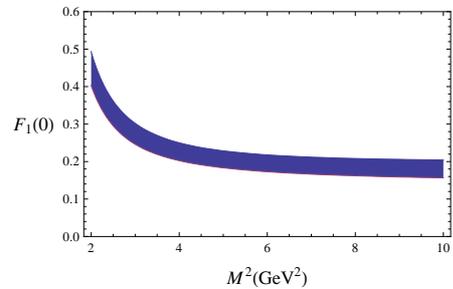}
\caption{Dependence of $F_1^{B_s \to f_0}(0)$ on the Borel parameter $M^2$. }\label{fig:M2dependence}\end{center}
\end{figure}
In Fig.~\ref{fig:M2dependence}  we show the dependence of  $F_1(q^2=0)$ on
$M^2$. We  observe stability when
$M^2>6$ ${\rm GeV}^2$, and we fix $M^2=(8\pm2)\,  {\rm GeV}^2$.

\begin{figure}[t]
\begin{center}
\includegraphics[width=0.35\textwidth]{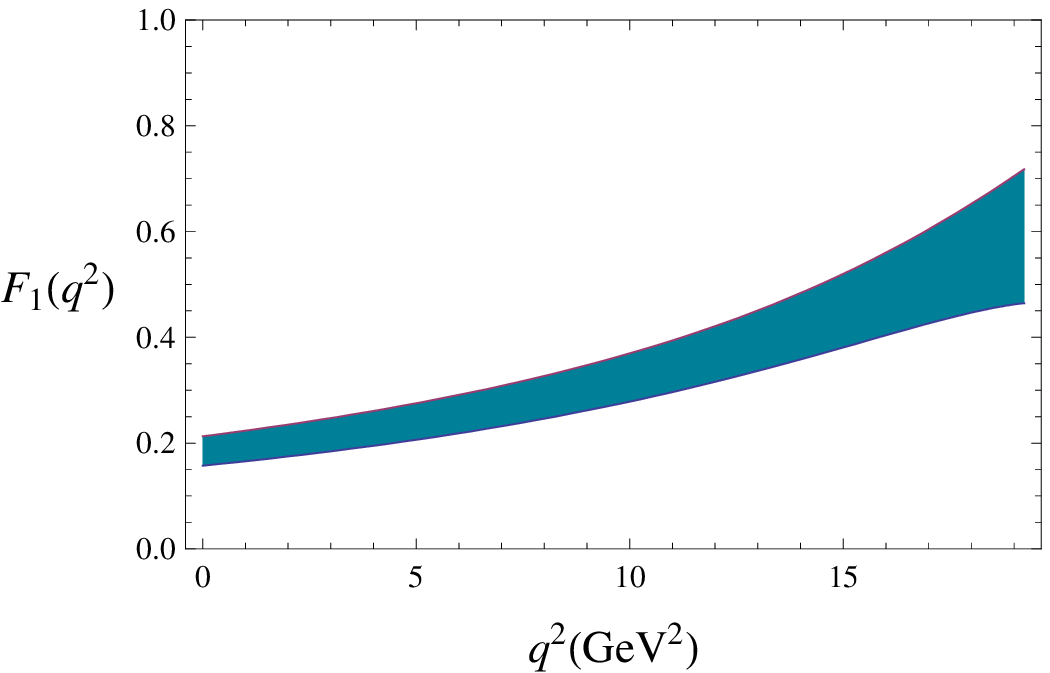}
\includegraphics[width=0.35\textwidth]{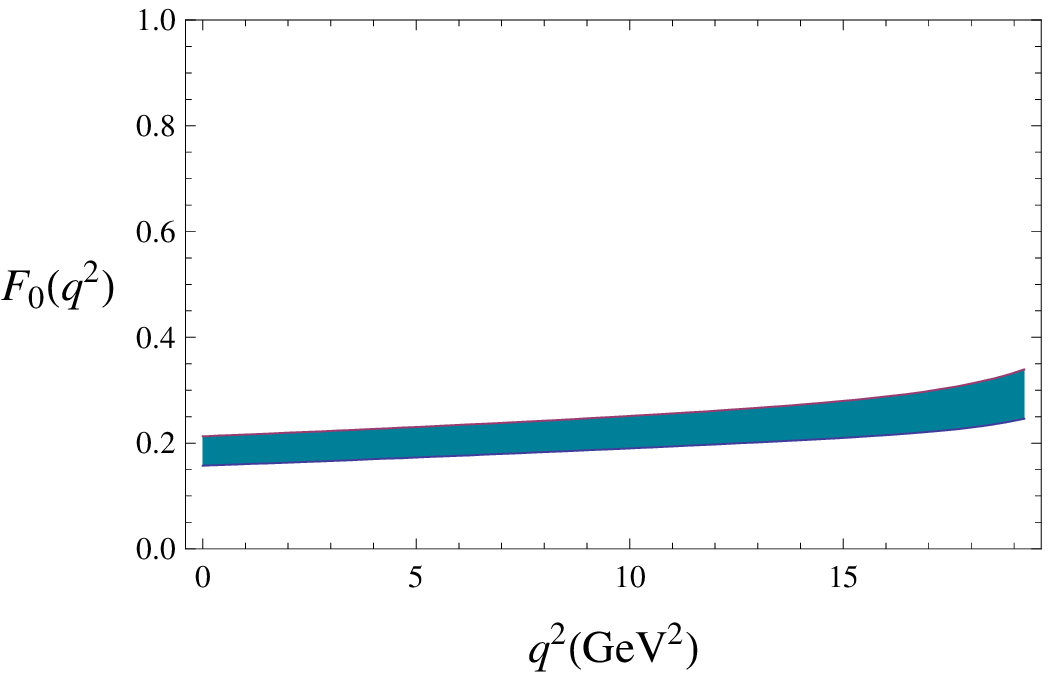}\\
\includegraphics[width=0.35\textwidth]{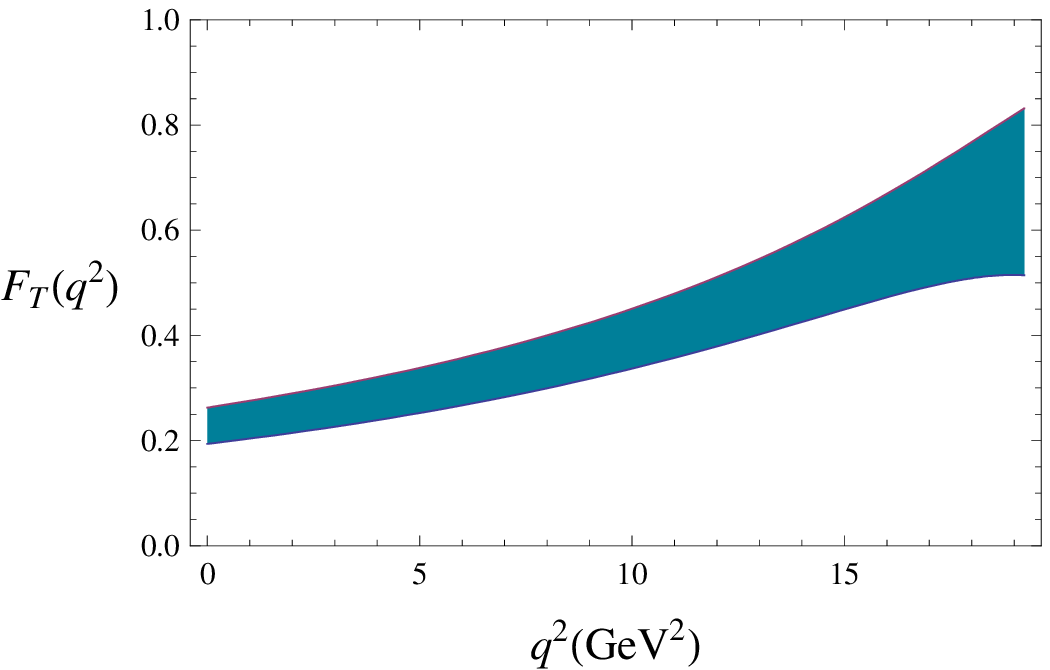}
 \caption{$q^2$ dependence of the $B_s\to
f_0$ form factors. }\label{fig:LO-formfactor}\end{center}
\end{figure}

To describe the form factors in the whole kinematically accessible
$q^2$ region, we use the  parameterization
$\displaystyle{F_i(q^2)=\frac{F_i(0)}{1-a_iq^2/m_{B_s}^2+b_i(q^2/m_{B_s}^2)^2}}$, $ i \in \{1,0,T\}$.
We collect in
Table~\ref{table:LO-formfactor} the
parameters $F_i(0)$,  $a_i$ and $b_i$  obtained  fitting the form
factors computed numerically. The  $q^2$ dependence is shown in Fig.~\ref{fig:LO-formfactor}.
 The uncertainties in the results are due to the input parameters,  $s_0$ and  $M^2$.
The  parameters $a_i$ and $b_i$
 are close for  $F_1$ and $F_T$. The reason is the following.
 In the heavy-quark limit and in the large energy  (LE) limit of the recoiled meson, the
 $B_s\to f_0$ form factors can be related \cite{Charles:1998dr} as follows:
\begin{eqnarray}
 \frac{m_{B_s}}{m_{B_s}+m_{f_0}}F_T(q^2)=F_1(q^2)=\frac{m_{B_s}}{2E}
 F_0(q^2),\label{eq:large-energy-limit}
\end{eqnarray}
with $q^2=m_{B_s}^2-2m_{B_s} E$ (neglecting $m_{f_0}^2$).
The first equality in (\ref{eq:large-energy-limit})  predicts the same $q^2$ dependence for
$F_1$ and $F_T$ in the LE limit. For the
parameters of $F_0$,
the second equality gives:
$a_0 =-1+a_1$, $b_0 = 1-a_1+b_1$ which, using the results for $a_1$ and $b_1$, gives
$a_0^{(LE)}\simeq 0.44 \pm 0.1$ and $b_0^{(LE)}\simeq 0.15 \pm
0.12$. Hence
 the first relation is
respected in our calculation,  while  not much can be said about the
second one due to the  uncertainty affecting  $b_0$.
\begin{table}
\caption{Parameters of the $B_s \to f_0$  form factors by LCSR. }\label{table:LO-formfactor}
\begin{center}
\begin{tabular}{ c c c c c }
\hline & $F_i(q^2=0)$  & $a_i$ & $b_i$
\\\hline
 $F_1 $   &   $0.185\pm0.029$
   &   $1.44^{+0.13}_{-0.09}$  &   $0.59^{+ 0.07}_{-0.05}$ \\
 $F_0 $   &   $0.185\pm0.029$
    &   $0.47^{+0.12}_{-0.09}$  &   $0.01^{+  0.08}_{-0.09}$\\
 $F_T $   &   $0.228\pm0.036$
    &   $1.42^{+0.13}_{-0.10}$  &   $0.60^{+  0.06}_{-0.05}$ \\
\hline
\end{tabular}\end{center}
\end{table}

\section{Semileptonic $\bar B_s\to f_0 \ell^+ \ell^-$ and $\bar B_s\to f_0
\nu\bar\nu$ decays}\label{PA-A}
\nin
$B_s$ decays induced by the   $b \to s$ transition can constrain new Physics scenarios. For example, they are sensitive to the compactification radius of universal extra dimensions \cite{Colangelo:2007jy}. Among such modes we consider $\bar B_s\to f_0\ell^+\ell^-$ and $\bar B_s\to f_0 \nu\bar\nu$, using the
 $B_s \to f_0$ form factors  to compute their branching ratios.

\nin The  SM effective Hamiltonian describing the  transition $b \to s \ell^+
\ell^-$ is:
\begin{equation}
H_{b \to s \ell^+ \ell^-}\,=-\,4\,{G_F \over \sqrt{2}} V_{tb}
V_{ts}^* \sum_{i=1}^{10} C_i(\mu) O_i(\mu) \,\,\,\, , \label{hamil}
\end{equation}
\noindent  $G_F$  being the Fermi constant and $V_{ij}$ the
elements of the CKM mixing matrix
(we neglect terms proportional to $V_{ub} V_{us}^*$). The expression of the operators $O_i$ can be found e.g in \cite{buchalla}.  The
Wilson coefficients  in (\ref{hamil}) are known
at NNLO in the SM  \cite{nnlo}.    $C_3-C_6$ are small,  hence  the contribution of only $O_7$, $O_9$ and $O_{10}$
can be kept for the description of the  $b \to s \ell^+  \ell^-$ transition. We use  a modified   $C_7^{eff}$, which is a renormalization
scheme independent combination of $C_7, C_8$ and $C_2$, given by a
 formula that  can be found, e.g., in \cite{Colangelo:2006vm}.

The   matrix elements of the operators in $H_{b \to s \ell^+ \ell^-}$ can be written in terms of form factors, so that   the differential decay width of $\bar B_s\to f_0\ell^+\ell^-$ reads:
\begin{eqnarray}
 \hspace{-0.7cm}\frac{d\Gamma(\bar B_s\to f_0\ell^+\ell^-)}{dq^2}=  \frac{G_F^2 \alpha^2_{em} |V_{tb}|^2|V^*_{ts}|^2}{512 m_{B_s}^3\pi^5}
 \sqrt{\frac{q^2-4m_\ell^2}{q^2}}\frac{ \sqrt{\lambda}}{3q^2} \nn \\ \hspace{-0.7cm}\Bigg\{ |C_{10}|^2 \Big[ 6m_\ell^2 (m_{B_s}^2-m_{f_0}^2)^2F_0^2(q^2)+(q^2-4m_\ell^2)\lambda F_1^2(q^2) \Big] \nonumber\\
\hspace{-0.7cm} +(q^2+2m_\ell^2)\lambda\bigg| C_9F_1(q^2)+\frac{2C_7^{eff}(m_b-m_s) F_T(q^2)}{m_{B_s}+m_{f_0}}\bigg|^2 \Bigg\},\nn
 \end{eqnarray}
 with
$\lambda=\lambda(m_{B_s}^2,m_{f_0}^2,q^2)=(m_{B_s}^2-q^2-m_{f_0}^2)^2-4m_{f_0}^2q^2$,
$\alpha_{em}$  the fine structure constant and $m_\ell$
the lepton mass.

\noindent Analogously, the  effective Hamiltonian  for  $b \to s \nu \bar \nu$ is
\begin{equation}
\hspace{-0.7cm} H_{b \to s\nu \bar \nu}= {G_F \over \sqrt{2}} {\alpha_{em}  V_{tb} V_{ts}^* \over 2 \pi
\sin^2(\theta_W)}\eta_X X(x_t) \, O_L  \equiv C_L
O_L \,\,, \label{hamilnu}
\end{equation}
 where
$O_L = \left( {\bar s}\gamma^\mu (1-\gamma_5) b \right) \left({\bar \nu}\gamma_\mu
(1-\gamma_5) \nu\right)$ and
 $\theta_W$  is the Weinberg angle; the  function $X(x_t)$
($x_t= m_t^2/ m_W^2$,  with $m_t$  the top
 mass and $m_W$ the $W$ mass) has been computed in \cite{inami}  and
\cite{buchalla,urban}, while   $\eta_X \simeq 1 $
\cite{buchalla,urban,Buchalla:1998ba}.
From $H_{b \to s\nu \bar \nu}$   the differential decay width is obtained:
\begin{eqnarray}
 \hspace{-0.6cm} \frac{d\Gamma(\bar B_s\to f_0\nu\bar\nu)}{dq^2}=3
 \frac{|C_L|^2\lambda^{3/2}(m_{B_s}^2,m_{f_0}^2,q^2)}{96 m_{B_s}^3 \pi^3}|F_1(q^2)|^2\,\,. \nonumber
\end{eqnarray}
\nin
Referring  to \cite{Colangelo:2010bg} for the values of the  parameters, we get:
\begin{eqnarray}
 {\cal BR}(\bar B_s\to f_0\ell^+\ell^-)&=&  (9.5^{+3.1}_{-2.6})\times
 10^{-8} \nn \\
 {\cal BR}(\bar B_s\to f_0\tau^+\tau^-)&=&(1.1^{+0.4}_{-0.3})\times 10^{-8}\\
 {\cal BR}(\bar B_s\to f_0\nu\bar\nu)&=&(
  8.7^{+2.8}_{-2.4})\times 10^{-7} \,\,\,  \nn
\end{eqnarray}
with $\ell=e,\mu$.
Hence these decays are  accessible at the  LHCb experiment at the CERN Large Hadron Collider
and at a  Super B factory operating at the $\Upsilon(5S)$ peak.
\section{Nonleptonic $B_s\to J/\psi f_0$ transition}\label{PA-B}
\nin
In the $B_s$ sector,   $B_s \to J/\psi \phi$ is  the golden mode to investigate CP violation. Analysing it,   CDF \cite{Aaltonen:2007he} and D0 \cite{:2008fj} Collaborations   have  obtained values of the $B_s$ mixing phase $\phi_s=-2\beta_s$ much larger than expected in the SM,  modulo a  large experimental uncertainty.  Hence, it is of prime importance  to consider other processes  to measure  $\beta_s$, as  $B_s \to J/\psi \eta^{(\prime)}$ and $ J/\psi f_0(980)$ in which   the final state is a CP eigenstate  and no angular analysis is required to disentangle  the various CP components,  as needed for  $B_s \to J/\psi \phi$.
However, the reconstruction of $B_s$ modes into $\eta$ and $\eta^\prime$ is experimentally challenging, since the subsequent $\eta$ or $\eta^\prime$ decays involve photons in the final state. The case of $f_0$ seems  feasible, since $f_0$ essentially decays to $\pi^+ \pi^-$ and to $2 \pi^0$ \cite{Amsler:2008zzb}.

From the theory viewpoint, the quantitative description of nonleptonic decays is  challenging. Using the operator product expansion  and renormalization group methods one can write an effective hamiltonian as for the modes in the previous section. However, now one has to consider hadronic matrix elements  $\langle J/\psi f_0 |O_i| B_s \rangle$ with $O_i$ four-quark operators, the calculation of which is a nontrivial task.
In order to estimate the size of the $B_s \to f_0 J/\psi$ decay rate, we  use the generalized factorization approach, in which such  quantities are replaced by products of matrix elements  that   are expressed in terms of  meson decay constants and hadronic form factors. The Wilson coefficients (or appropriate combinations of them) are regarded as effective parameters to be fixed from experiment.
Using this  ansatz, the decay amplitude of $\bar B_s\to J/\psi f_0$
reads
\begin{eqnarray}
\hspace{-0.7cm} A(\bar B_s\to J/\psi f_0)= \frac{2G_F}{\sqrt 2}V_{cb}V_{cs}^* a_2 m_\psi
 f_{J/\psi}F_1(m_{J/\psi}^2) (\epsilon^*  p_{B_s})\nn
\end{eqnarray}
where $\epsilon$ is the $J/\psi$ polarization vector, $p_{B_s}$ the $B_s$ momentum and
 $ f_{J/\psi}=(416.3\pm5.3)$ MeV  the  $J/\psi$ decay constant.
$a_2$ is a combination of Wilson coefficients that
 can be extracted from   ${\cal BR}(B\to J/\psi K)$   \cite{Amsler:2008zzb},  assuming  that $a_2$ is the same in the two processes.
This requires the form factor $F_1^{B \to K}$.
We use two different parameterizations, obtained by  short-distance (CDSS) \cite{Colangelo:1995jv}  and light-cone QCD sum rules (BZ) \cite{Ball:2004ye}. The result  for the two sets of form factors is:
$
 |a_2^{ (CDSS)}|=0.394^{+0.053}_{-0.041}$, $
 |a_2^{ (BZ)}|=0.25\pm 0.03$.
We use
the average value $a_2=0.32\pm 0.11$ and our result for  the $B_s \to f_0$ form factors to  compute
${\cal BR}(\bar B_s\to J/\psi f_0)$,  obtaining
\begin{eqnarray}
 {\cal BR}(\bar B_s\to J/\psi f_0)=(3.1\pm 2.4)\times
 10^{-4}\label{eq:Bs-Jpsi-f0-LO}\,,
\end{eqnarray}
large enough to be measured; notice that
$ {\cal BR}(B_s\to J/\psi \phi)=(1.3\pm 0.4)\times10^{-3}$ \cite{Amsler:2008zzb}.
Comparing these results to  ${\cal BR}(B_s\to J/\psi_L\phi_L)$ ($L$ denotes a longitudinally
polarized meson) computed using factorization, we find:
\begin{eqnarray}
R_{f_0/\phi}^{B_s}= \frac{{\cal BR}(B_s\to J/\psi f_0)}{{\cal BR}(B_s\to
 J/\psi_L\phi_L)} =0.13\pm 0.06  \,. \label{Rphi}
\end{eqnarray}
A compatible result: $R_{f_0/\phi}^{B_s}\simeq 0.2-0.3$ was found in \cite{Stone:2008ak}, using the  ratio of $D_s$ decay widths to $f_0 \pi^+$ and $\phi \pi^+$.

These  considerations show that  $B_s \to J/\psi f_0$ can be used to measure $\beta_s$, since  a large number of events is expected and it
does not require an angular analysis to separate  different CP components of the final state.  This is also  the case of   $B_s\to \chi_{c0}\phi$, modulo the  difficulty of the $\chi_{c0}$ reconstruction.
Although  suppressed  in naive factorization,  its
branching fraction may  be enhanced by non factorizable  mechanisms~\cite{Colangelo:2002mj}  as for $B\to\chi_{c0}K$.  On the basis of  $SU(3)_F$ symmetry, we expect ${\cal BR}(B_s\to \chi_{c0}\phi)\simeq {\cal O}(10^{-4})$ as in the case of
 $B\to \chi_{c0}K^*$ \cite{:2008hj}.
\section{Conclusions}
\nin
Exploiting the LCSR calculation of $B_s\to f_0$  form factors we find that the branching ratios of   $B_s\to
f_0\ell^+\ell^-$ and $B_s\to f_0\nu\bar\nu$ will be  accessible at   future machines, like a Super B
factory,  and  at the LHCb experiment.
We also predict ${\cal BR}(B_s\to J/\psi f_0)/{\cal BR}(B_s\to
J/\psi \phi)=0.13 \pm 0.06$,  thus $B_s\to J/\psi
f_0$ is  promising to access $\beta_s$.
\section*{Acknowledgements}
\nin
I thank P. Colangelo and W. Wang for collaboration and M. Nielsen for discussions. I acknowledge  the RTN  FLAVIAnet MRTN-CT-2006-035482 (EU) for support.








\end{document}